\PassOptionsToPackage{table,dvipsnames}{xcolor}   
\documentclass[sigconf,nonacm]{acmart}
\usepackage{pifont}
\usepackage[mathcal]{euscript}

\usepackage{placeins}

\usepackage{float}
\usepackage{hhline}
\usepackage{makecell}
\usepackage[most]{tcolorbox}
\usepackage{threeparttable}

\usepackage{subcaption}
\usepackage{tikz}
\usepackage{standalone}
\usetikzlibrary{arrows.meta,shapes.geometric}

\usepackage[ruled,vlined]{algorithm2e}
\usepackage{algpseudocode}

\usepackage{xurl}

\usepackage{enumitem}

\usepackage[
  separate-uncertainty = true,
  multi-part-units = repeat
]{siunitx}

\usepackage{adjustbox}
\usepackage{array}
\usepackage{longtable}
\usepackage{tabularx}
\usepackage{ragged2e}
\newcolumntype{Y}{>{\raggedright\arraybackslash}X}
\usepackage{arydshln}
\usepackage{multirow}

\newtcolorbox{GrayBox}[2][]{%
  enhanced,
  float=htbp,
  width=\columnwidth,
  colback=gray!10,
  colframe=black!70,
  arc=1mm,
  boxrule=0.5pt,
  left=1ex,
  right=1ex,
  top=0.5ex,
  bottom=0.5ex,
  before skip=4pt,
  after skip=4pt,
  title=#2,
  #1
}

\newcommand{\parg}[1]{%
  \par\medskip \noindent\textbf{\textit{#1}}\quad%
}

\newcommand{\best}[1]{\cellcolor{gray!25}{#1}}

\definecolor{gainbase}{HTML}{A6CEE3}
\definecolor{lossbase}{HTML}{CAB2D6}
\newcommand{\gc}[1]{\cellcolor{gainbase!#1!white}}
\newcommand{\rc}[1]{\cellcolor{lossbase!#1!white}}

\newcommand{\zs}{UXC\textsubscript{ZS}}
\newcommand{\ft}{UXC\textsubscript{FT}}
\newcommand{\mlp}{$\mathcal{N}$\textsubscript{MLP}}

\newcommand{\paperabstract}{%
Code-to-architecture mapping assigns source files to modules of a reference architecture. Representations of implementation artifacts can capture code text, dependencies, and directory organization, but predictive gains alone do not establish the contribution of additional graph processing. Dependencies may already be encoded in these representations, while directories may closely match the reference decomposition. We investigate how artifact representations and graph context affect mapping on ten Java, C, and C++ systems with a limited labeling budget. We compare lexical, structural, and contextual file representations and assess each predictor with and without dependency neighbors. Using a Heterogeneous Graph Transformer (HGT), we compare file-level graphs extended with member nodes summarizing each file’s program elements, folder nodes, or both, alongside a variant encoding folder paths directly in file representations. We compare folder-aware predictions with a directory rule using the same labeled files.

Representations encoding dependency information outperform lexical representations on most systems. Additional neighbor aggregation can improve mapping, depending on the predictor and aggregation direction. Folder-aware variants yield the largest and most consistent gains across the evaluated HGT configurations. However, the directory rule achieves per-system median accuracy of 89–100\%, and adding folder information does not consistently improve predictions for files the rule misclassifies. Member nodes provide no consistent improvement over file-level graphs. These findings support assessing what file representations and directory organization already provide before attributing improvements to graph neural networks.
}

\title{What Carries Architectural Signal? A Code-to-Architecture Mapping Study}

\author{Rakhshanda Jabeen}
\affiliation{\institution{Electrolux Professional AB}\city{Ljungby}\country{Sweden}}
\affiliation{\institution{Linnaeus University}\city{V\"axj\"o}\country{Sweden}}
\email{rakhshanda.jabeen@electroluxprofessional.com}

\author{Morgan Ericsson}
\affiliation{\institution{Linnaeus University}\city{V\"axj\"o}\country{Sweden}}
\email{morgan.ericsson@lnu.se}

\author{Yuyao Duan}
\affiliation{\institution{Linnaeus University}\city{V\"axj\"o}\country{Sweden}}
\email{yuyaoduan4@gmail.com}

\author{Anna Wingkvist}
\affiliation{\institution{Linnaeus University}\city{V\"axj\"o}\country{Sweden}}
\email{anna.wingkvist@lnu.se}

\begin{abstract}
\paperabstract
\end{abstract}


\begin{document}
\makeatletter
\renewcommand\subsubsection{%
  \@startsection{subsubsection}{3}{\z@}%
    {-1.5ex plus -0.5ex minus -0.2ex}%
    {0.5ex plus 0.2ex}%
    {\normalfont\normalsize\itshape}%
}
\let\ACM@origsubsubsection\subsubsection
\makeatother

\raggedbottom
\maketitle

\section{Introduction\label{sec:intro}}

Maintaining a software architecture requires knowing which implementation elements belong to its modules~\cite{parnas1972,clements2003}. This correspondence can become incomplete as a system grows and architectural documentation falls behind the implementation. Code-to-architecture (C2A) mapping addresses this problem by assigning source files to the modules of a reference architecture. When only part of a system has been mapped, automated approaches can use these known assignments together with information from the code and project structure to infer the remaining assignments.

Automated C2A mapping can draw on several sources of evidence. Identifiers express domain concepts and implementation roles, while declarations and operations describe how those responsibilities are implemented. Static dependencies describe interactions between files, and directories record how developers organize a project~\cite{sarif,negar,garcia2013}. These sources need not agree: a utility may interact with several modules, terminology may be shared across responsibilities, and a directory may contain files assigned to different architectural modules. This study examines what dependency-neighbor aggregation and richer graph structures add beyond the information available in file representations and directory organization.

Prior C2A mapping work combines lexical features with static dependencies in graph neural networks~\cite{jabeen2025graph}, with later work also incorporating folder organization~\cite{c2a_mapping}. Related work on architecture recovery, which infers a decomposition without a reference architecture, explores random-walk and graph-autoencoder representations of dependencies~\cite{negar,structsar} and encodings of identifier and folder tokens~\cite{gaer}. When several information sources and predictor mechanisms change together, comparisons of overall performance leave the contribution of additional graph context unclear.

A classifier may already receive dependency information through its input vectors, so the contribution of additional neighbor aggregation needs to be evaluated separately. Similarly, high agreement with a reference architecture may partly reflect agreement between that architecture and the directory structure. A folder-aware model can improve overall accuracy while reducing accuracy on files that a simple directory rule misclassifies. Aggregate scores alone do not distinguish these outcomes. Representing members separately may expose context obscured in a single file-level representation.

We investigate these questions on ten Java, C, and C++ systems, using a bounded labeling budget with at least one labeled file from every module. The setting is transductive: features and graph structure are available for labeled and unlabeled files, while module-label supervision uses only the initially labeled files. We compare lexical Word2Vec, structural Node2Vec, and contextual UniXcoder representations. A multilayer perceptron (MLP) establishes their performance before additional neighbor aggregation. We use a Heterogeneous Graph Transformer (HGT)~\cite{hgt} to compare file, folder, and member configurations, and assess folder-aware predictions against a directory rule using the same labeled files. We also compare performance with neighbors enabled and disabled for HGT, a neighbor MLP, and a Graph Attention Network (GAT)~\cite{gat}, with aggregation in one or both directions.

\noindent The study is guided by three research questions:
\begin{itemize}[leftmargin=1em, itemsep=1pt]
    \item \textbf{RQ1.} How does mapping performance vary across lexical, structural and contextual file representations?
    \item \textbf{RQ2.} How does adding member and folder nodes to file-level HGT change mapping performance, and how do folder-aware predictions compare with a directory rule?
    \item \textbf{RQ3.} How does using dependency neighbors change mapping performance, and how do the predictor and aggregation direction affect this change?
\end{itemize}

The results show that file representations encoding dependency information provide strong starting points, while the benefit of further neighbor aggregation depends on the predictor and aggregation direction. Folder nodes yield the largest gains within the HGT configurations, but the directory rule reaches similar accuracy on most systems, and folder paths supplied directly as file features match or exceed folder nodes on most systems. These overall gains do not consistently extend to files the rule assigns incorrectly. Member nodes provide no consistent improvement with one or two HGT layers. Code and data are available on GitHub\footnote{\label{fn:code}\url{https://github.com/sse-lnu/hetmap}}.

\section{Methodology\label{sec:method}}

This section presents the methodology for studying how architectural signals and graph design choices affect C2A mapping. We first formulate the mapping task and then describe the graphs and the structural, lexical, and contextual features used to initialize their nodes. We then present the MLP, HGT, and GAT predictors.

Each research question varies one of these components and holds the others fixed. RQ1 varies the file representation, using the MLP. RQ2 varies the graph variant for each representation, using HGT because it handles the node and relation types of the heterogeneous variants. RQ3 varies the predictor, comparing the neighbor MLP with GAT and HGT, and the dependency neighbors they receive.

\subsection{Problem Formulation\label{sec:problem}}

We formulate the mapping task as node classification over a software graph $G=(V, E,\mathcal{A},\mathcal{R})$, where $V$ and $E$ are the sets of artifact nodes and relations between them, respectively, and $\mathcal{A}$ and $\mathcal{R}$ denote the sets of node and relation types. The mappings $\tau: V \rightarrow \mathcal{A}$ and $\phi: E \rightarrow \mathcal{R}$ assign node and relation types. Let $V_f \subseteq V$ be the set of file nodes, which are the targets of the mapping task. A small subset $L \subset V_f$ is labeled with known architectural module assignments $Y_L$, and the remaining file nodes $U = V_f \setminus L$ are unlabeled. The task is to learn a model:

\begin{equation}
    f_{\theta}: (G, L, Y_L) \rightarrow \hat{Y}_U,
\end{equation}

where $\theta$ denotes the learnable parameters. The model is trained on the labeled files only, but all nodes of $G$, labeled and unlabeled, are part of the graph during training.

Each model computes a $d$-dimensional embedding $z_i$ for every file $i \in V_f$, and the architectural module of a file is predicted from this embedding.

\subsection{Graph Construction\label{sec:het_graph}\label{sec:graph_variants}}

We represent the software system as a graph of software artifacts connected by folder organization, member containment, and static dependencies.

\begin{figure}[h]
\centering
\resizebox{0.97\linewidth}{!}{\begin{tikzpicture}[
  x=1cm,y=1cm,
  font=\sffamily\normalsize, 
  folder/.style={
    rectangle,
    draw=black!70,
    fill=gray!10,
    line width=.75pt,
    minimum width=5mm,
    minimum height=3.5mm,
    inner sep=0pt
  },
  file/.style={
    circle,
    draw=black!70,
    fill=gray!10,
    line width=.75pt,
    minimum size=6mm,
    inner sep=0pt
  },
  member/.style={
    diamond,
    aspect=1,
    draw=black!70,
    fill=gray!10,
    line width=.75pt,
    minimum size=7mm,
    inner sep=.5pt
  },
  edge/.style={
    -{Stealth[length=1.7mm,width=1.2mm]},
    draw=black!85,
    line width=.7pt,
    line cap=round,
    line join=round,
    shorten <=1.2pt,
    shorten >=1.5pt
  },
  relation/.style={
    edge,
    dash pattern=on 2.5pt off 1.7pt
  },
  dependency/.style={edge},
  panel title/.style={
    font=\fontsize{10}{12}\selectfont
  },
  rel label/.style={
    font=\fontsize{9}{10}\selectfont\itshape,
    text=black,
    inner sep=1pt
  },
  legend text/.style={
    anchor=west,
    inner sep=0pt,
    align=left
  }
]

\begin{scope}
  \draw[
    draw=gray!20,
    line width=.4pt,
    rounded corners=3pt
  ] (-.15,1.05) rectangle (2.55,5.2);

  \node[folder] at (.3,4.8) {};
  \node[legend text] at (.7,4.8) {Folder};

  \node[file,minimum size=5mm] at (.3,4.05) {};
  \node[legend text] at (.7,4.05) {File};

  \node[member,minimum size=6mm] at (.3,3.3) {};
  \node[legend text] at (.7,3.3) {Member};

  \draw[relation] (0,2.4) -- (.6,2.4);
  \node[legend text] at (.7,2.4) {Structural};

  \draw[dependency] (0,1.5) -- (.6,1.5);
  \node[legend text] at (.7,1.5) {Dependency};
\end{scope}

\begin{scope}[xshift=2.8cm]

  \begin{scope}[shift={(0,6)}]
    \node[panel title] at (1,.75) {\normalsize F};

    \node[file] (f1) at (0,0) {$f_1$};
    \node[file] (f2) at (2,0) {$f_2$};

    \draw[dependency] (f1) -- (f2);
  \end{scope}

  \begin{scope}[shift={(4,6)}]
    \node[panel title] at (1,.75) {\normalsize FM};

    \node[file] (fmf1) at (0,0) {$f_1$};
    \node[file] (fmf2) at (2,0) {$f_2$};
    \node[member] (fmm1) at (0,-1.3) {$m_1$};
    \node[member] (fmm2) at (2,-1.3) {$m_2$};

    \draw[dependency] (fmf1) -- (fmf2);
    \draw[dependency] (fmm1) -- (fmm2);
    \draw[relation] (fmm1) -- (fmf1);
    \draw[relation] (fmm2) -- (fmf2);
  \end{scope}

  \begin{scope}
    \node[panel title] at (1,3.65) {\normalsize FF};

    \node[folder] (ffp) at (1,2.9) {};
    \node[folder] (ffd) at (1,1.6) {};
    \node[file] (fff1) at (0,0) {$f_1$};
    \node[file] (fff2) at (2,0) {$f_2$};

    \draw[relation] (ffp) -- (ffd);
    \draw[relation] (ffd) -- (fff1);
    \draw[relation] (ffd) -- (fff2);
    \draw[dependency] (fff1) -- (fff2);
  \end{scope}

  \begin{scope}[xshift=4cm]
    \node[panel title] at (1,3.65) {\normalsize FFM};

    \node[folder] (ffmp) at (1,2.9) {};
    \node[folder] (ffmd) at (1,1.6) {};
    \node[file] (ffmf1) at (0,0) {$f_1$};
    \node[file] (ffmf2) at (2,0) {$f_2$};
    \node[member] (ffmm1) at (0,-1.3) {$m_1$};
    \node[member] (ffmm2) at (2,-1.3) {$m_2$};

    \draw[relation] (ffmp) -- (ffmd)
      node[rel label,midway,left=3pt] {parent\_of};

    \draw[relation] (ffmd) -- (ffmf1)
      node[rel label,midway,left=3pt] {contains};

    \draw[relation] (ffmd) -- (ffmf2);

    \draw[dependency] (ffmf1) -- (ffmf2)
      node[rel label,midway,below=4pt] {depends\_on};

    \draw[dependency] (ffmm1) -- (ffmm2)
      node[rel label,midway,below=4pt] {depends\_on};

    \draw[relation] (ffmm1) -- (ffmf1)
      node[rel label,midway,left=3pt] {belongs\_to};

    \draw[relation] (ffmm2) -- (ffmf2);
  \end{scope}

\end{scope}

\end{tikzpicture}}
\caption{Graph variants F, FM, FF, and FFM. Files $f_i$ are the prediction targets, and each $m_i$ aggregates the members of file $f_i$. Arrows point from source to target: dashed arrows are structural relations and solid arrows are dependencies. Relation names are shown on FFM only. The panels omit further dependency edges and folder levels; Table~\ref{tab:relation_types} lists the relations with the node types they connect. Feature initialization also depends on the variant, as described in Section~\ref{sec:node_features}.}
\label{fig:graph_variants_overview}
\end{figure}
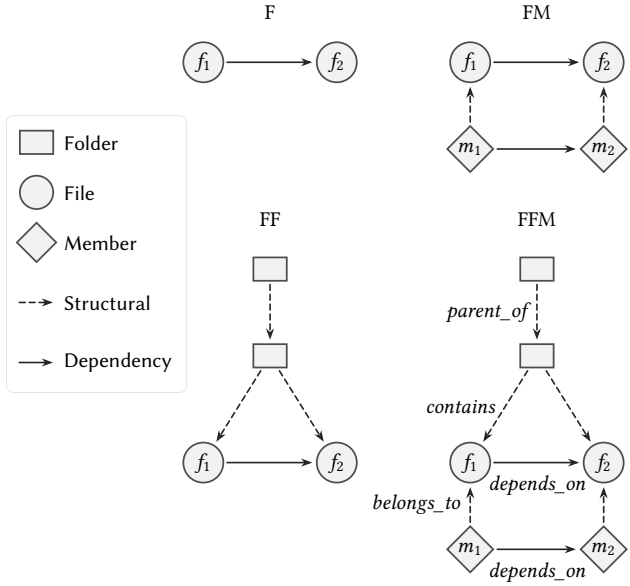

To study the individual and combined contributions of folder and member information, we construct four graph variants that differ in the node types they contain. They are named after these types: F (file nodes only), FF (folder and file nodes), FM (file and member nodes), and FFM (folder, file, and member nodes). All variants retain the dependency relation. F is homogeneous, with one node type and one relation type, and the other three variants are heterogeneous. Figure~\ref{fig:graph_variants_overview} provides a visual overview.

The complete graph (FFM) has three node types. File nodes represent source files and are the targets of C2A mapping. Because logical components are often implemented through namespaces or directory structures that group source code providing a particular functionality~\cite{fundamentals_sa, garcia2013}, folder nodes represent directory levels of the project, with one node per level. Each file has one summarized member node covering the methods, fields, constructors, and variables of that file. This retains member-level information while avoiding the large and sparse graphs produced by representing individual members.

Four relations connect these nodes, as listed in Table~\ref{tab:relation_types}. Three are structural and encode organization and containment, and the fourth encodes static dependencies. Folder nodes are connected through \emph{parent\_of} edges, forming chains such as \texttt{app} $\rightarrow$ \texttt{app/ui} $\rightarrow$ \texttt{app/ui/views}, and \emph{contains} edges link each folder to its files. A \emph{belongs\_to} edge links each member node to its enclosing file.

Messages along structural relations flow toward the file. Folder information passes down the hierarchy through \emph{parent\_of} and \emph{contains} edges, and member information passes up to the file through \emph{belongs\_to} edges. No reverse edges are added, so folder and member nodes are not updated from files through structural relations.

\emph{depends\_on} encodes static dependencies, which provide direct evidence of how source files interact~\cite{negar, deps_impact}. These dependencies were extracted with the Depends tool~\cite{depends}, which analyzes the syntactic structure of the projects and identifies entities together with several categories of static relations. We merge all dependency types into a single relation, drop duplicate edges, and remove edges whose endpoints belong to the same file.

Dependency endpoints depend on the graph variant. In FM and FFM, dependencies connect either two files or two member nodes. Dependencies between a file and a member are resolved between their corresponding files. In F and FF, all member endpoints are replaced by their enclosing files, so all dependencies connect file nodes.

RQ1 uses no graph structure, and RQ3 uses the homogeneous file-level graph F. RQ2 uses all four variants to examine progressively richer graph structure, adding folder nodes in FF, member nodes in FM, and both in FFM.

\begin{table}[t]
\centering
\caption{Graph relations and node types.}
\label{tab:relation_types}
\renewcommand{\arraystretch}{1.1}
\begin{tabularx}{\columnwidth}{@{}l l >{\raggedright\arraybackslash}X@{}}
\toprule
Relation & Source $\to$ target & Description \\
\midrule
\emph{parent\_of}
& folder $\to$ folder
& Directory hierarchy. \\
\emph{contains}
& folder $\to$ file
& Folder contains a file. \\
\emph{belongs\_to}
& member $\to$ file
& Links a member node to its file. \\
\addlinespace[4pt]
\multirow{2}{*}{\emph{depends\_on}}
& file $\to$ file
& File-level dependencies. \\
& member $\to$ member
& Member-level dependency in FM and FFM. \\
\bottomrule
\end{tabularx}
\end{table}

\subsection{Node Feature Initialization \label{sec:node_features}}

Folder nodes are initialized in the same way in every setting. File and member nodes are initialized with one of the representations compared in RQ1: lexical embeddings from Word2Vec, structural embeddings from Node2Vec, or contextual embeddings from UniXcoder, used zero-shot or fine-tuned with dependencies.

\subsubsection{Folder Node Features}

Folder nodes use a binary bag-of-words representation of the folder tokens along their paths, after removing common project-root tokens. For example, \texttt{app/ui/views} is represented by \texttt{app}, \texttt{ui}, and \texttt{views}. Including the tokens of its ancestors preserves each folder's hierarchical context.

\subsubsection{Word2Vec (W2V): Lexical Features\label{sec:node_w2v}}

Source code identifiers encode domain concepts and implementation roles, making them useful signals for C2A mapping~\cite{sarif}. Following Jabeen et al.~\citep{gaer}, we extract tokens from file names and from the identifiers of members, namely methods, fields, constructors, and variables. We split camelCase, PascalCase, and snake\_case identifiers, convert tokens to lowercase, and remove stopwords and non-alphabetic tokens. Each node vector is the mean of the skip-gram Word2Vec embeddings of its associated tokens~\cite{word2vec}.

In FM and FFM, file nodes use file-name tokens, while member nodes use member identifiers. In F and FF, member identifiers are included in the file vector instead, and the MLP uses this same file vector. Member identifiers therefore remain available in every variant.

\subsubsection{Node2Vec (N2V): Structural Features \label{sec:node_n2v}}

Chen et al.~\citep{negar} observe that files in the same architectural module can occupy a dense local dependency region without being directly connected. They use random-walk embeddings to capture this longer-range structure for architecture recovery. In our earlier study of dependency-only recovery, which follows their setup, Node2Vec was among the best-performing learned representations~\cite{structsar}. Therefore, we use Node2Vec~\cite{node2vec} embeddings computed with uniform random walks on the undirected file-level dependency graph as a structural initialization. These embeddings encode dependency structure without using identifiers or folder location.

In FM and FFM, each summarized member node receives the Node2Vec embedding of its corresponding file.

\subsubsection{UniXcoder: Contextual Features \label{sec:unixcoder}}

For contextual initialization, we use UniXcoder~\cite{unixcoder}, a pretrained code transformer. We parse each source file with Tree-sitter~\cite{tree_sitter} using the corresponding Java, C, or C++ grammar. The syntax tree determines which source fragments are passed to the encoder, but is not itself an input to the model. Extraction produces two views. The file-level view contains type declarations, fields or global variables, and constructor signatures with short body excerpts where applicable. The member-level view contains methods or functions. Each view produces one vector per file. Member-level text is split into overlapping chunks whose embeddings are mean-pooled.

The zero-shot variant (\zs{}) uses the pretrained weights. UniXcoder's pretraining objectives do not target architectural grouping: files with similar code can belong to different modules. We therefore also evaluate a dependency-guided variant (\ft{}), fine-tuned separately for each system using an InfoNCE objective~\cite{infonce}. This objective brings dependency-based positive pairs closer than other in-batch examples treated as negatives. However, files that are close in the dependency graph need not belong to the same architectural module. During fine-tuning, we compute this loss using only the file-level views, without module labels. We then use the fine-tuned encoder to produce representations for both file- and member-level views.

Positive pairs are drawn from the two-hop neighborhood to capture longer-range dependency structure~\cite{structsar}, with dependencies treated as undirected. For each anchor file, two positives are sampled at random from the eight two-hop candidates that share the most one-hop neighbors with it. Files without a two-hop candidate fall back to a direct neighbor. Other files in the batch serve as negatives, except directly connected files, which are masked unless selected as the positive pair.

Member information enters the model differently across representations. Word2Vec includes member identifiers in file vectors when member nodes are absent and places them in member nodes otherwise. UniXcoder provides member text only through member nodes, while Node2Vec assigns each member node its file's vector. Adding member nodes thus changes the placement or availability of information as well as the graph structure, so comparisons across graph variants are interpreted separately for each representation.

\subsection{Mapping Models\label{sec:mapping_models}}

We compare models that use file features alone with models that incorporate graph structure. All models use the same classification head to predict architectural modules. Throughout, $\mathcal{N}(\cdot)$ denotes the neighborhood of a node.

\subsubsection{Multilayer Perceptron (MLP)\label{sec:mlp}}

The MLP predicts module logits directly from a file-level input vector. Each hidden layer computes:

\begin{equation}
h_i^{(k+1)} = \text{ReLU}\big(W^{(k)} h_i^{(k)} + b^{(k)}\big),
\end{equation}

with $h_i^{(0)} = x_i$ the file feature vector and $W^{(k)}, b^{(k)}$ the learnable parameters of layer $k$, and ReLU is the rectified linear unit~\cite{relu}. The input $x_i$ contains the single representation being evaluated.

For RQ3, we incorporate dependency neighborhood information in the MLP by concatenating each file's vector $x_i$ with the mean $\bar{x}_{\mathcal{N}(i)}$ of its direct neighbors' vectors in the dependency graph, giving $[x_i \,\|\, \bar{x}_{\mathcal{N}(i)}]$. We refer to the MLP with this input as the neighbor MLP (\mlp{}). With neighbors disabled, the input is $x_i$ alone, as in RQ1, so enabling neighbors also doubles the input width of the first layer.

\subsubsection{Heterogeneous Graph Transformer (HGT)\label{sec:hgt_model}}

HGT~\cite{hgt} conditions attention and message transformation on the source node type, edge type, and target node type. We describe an edge $e_{s,t}$ from node $s$ to node $t$ by the meta-relation $\langle \tau(s), \phi(e_{s,t}), \tau(t) \rangle$. For example, $\langle \text{folder}, \text{contains}, \text{file} \rangle$ captures project organization, whereas $\langle \text{file}, \text{depends on}, \text{file} \rangle$ captures code-level coupling.

Node types have different feature spaces depending on the graph variant and initialization strategy. We first map the input vector $x_t$ of each node $t$ into a shared hidden space using a type-specific linear transformation followed by ReLU, which gives its initial representation $h_t^{(0)}$. One HGT layer then updates node representations through multi-head attention over their neighbors.

For each attention head, the layer computes a message $M(s,e,t)$ and an attention weight $\alpha(s,e,t)$ for each source node $s$ connected to a target node $t$ by an edge $e$. We omit attention-head indices for brevity. The weighted messages are aggregated as:

\begin{align}
\tilde{m}_t &= \sum_{s \in \mathcal{N}(t)}
       \alpha(s,e,t)\, M(s,e,t), \\
M(s,e,t) &= W^{\mathrm{MSG}}_{\phi(e)}\,
            V_{\tau(s)}\, h_s, \notag
\end{align}
\begingroup
\fontsize{9.5pt}{12pt}\selectfont
\begin{equation*}
\alpha(s,e,t)
= \operatorname*{softmax}_{s \in \mathcal{N}(t)}
\frac{
  \big(K_{\tau(s)} h_s\big)^{\top}
  W^{\mathrm{ATT}}_{\phi(e)}\,
  \big(Q_{\tau(t)} h_t\big)\,
  \mu_{\langle \tau(s),\phi(e),\tau(t)\rangle}
}{\sqrt{d_h}}.
\end{equation*}
\endgroup

where $\tau(\cdot)$ is the node type, $\phi(e)$ is the edge type, $V$, $K$, and $Q$ are type-specific linear transformations, $W^{\mathrm{MSG}}$ and $W^{\mathrm{ATT}}$ are edge-type-specific matrices, $\mu$ is a learned prior for each meta-relation, and $d_h$ is the head dimension. These parameters allow a \emph{contains} edge from a folder and a \emph{depends\_on} edge from a dependent file to receive different attention weights.

In our implementation, the attention-head outputs are combined into a message $m_t$. We add this message to the node's projected input $h_t^{(0)}$ through a residual connection~\cite{residual_connection}, followed by layer normalization:

\begin{equation}
h_t^{(1)} = \mathrm{LayerNorm}\left(m_t + h_t^{(0)}\right).
\end{equation}

\subsubsection{Graph Attention Network (GAT)\label{sec:gat_model}}

GAT~\cite{gat} aggregates neighbor representations using attention parameters shared across neighbors within each head, without the node-type- or edge-type-specific parameters of HGT. The outputs of the attention heads are concatenated. Our GAT implementation uses the same input projection, residual connection, and layer normalization as HGT, while replacing its neighbor aggregation mechanism. We add no self-loops; the residual connection retains the node's own representation.

\subsubsection{Classification Head\label{sec:head}}

All three predictors use the same classification head. The file embedding $z_i \in \mathbb{R}^{d}$, obtained from the last hidden layer of the MLP or the message-passing layer of GAT and HGT, is passed to a linear layer that produces one score per architectural module. A softmax converts these scores into a probability distribution:

\begin{equation}
\label{eq:head}
\begin{aligned}
p_i &= \text{softmax}\big(W_c z_i + b_c\big), \\
\hat{y}_i &= \arg\max_m p_{i,m},
\end{aligned}
\end{equation}

where $W_c \in \mathbb{R}^{|M| \times d}$, $b_c \in \mathbb{R}^{|M|}$, and $|M|$ is the number of architectural modules. The predicted module $\hat{y}_i$ is the one with the highest probability. Training minimizes categorical cross-entropy over the labeled file nodes.

\section{Experimental Setup}

This section describes the benchmark systems, the experimental configuration, the directory-based reference predictor, and the evaluation procedure used to answer the three research questions.

\subsection{Benchmark Systems\label{sec:benchmark_systems}}

We evaluate the models on ten open-source systems with published ground-truth (GT) mappings, which have been used in prior work on architecture recovery and C2A mapping~\cite{gaer,olsson2022automatically}. They differ in programming language, domain, size, and architectural granularity. We use published mappings because reliable C2A mappings require substantial manual effort and expert validation~\cite{garcia2013}, and because they allow comparison with previous studies.

Table~\ref{tab:datasets} summarizes the system and graph properties. File-level dependency homophily $\mathcal{H}$ is the proportion of unique directed file pairs connected by a static dependency whose source and target files fall within the same GT module, from 0 (every dependency crosses a module boundary) to 1 (every dependency stays within a module). The mean degree $\langle k \rangle = 2|E|/|\text{Files}|$ counts each of the $|E|$ directed file-level dependency edges at both of its files.

\begin{table}[h]
\centering
\caption{Subject systems and their file-level dependency graph properties. \(M\): number of modules in the GT; \(|E|\): number of directed file-level static dependencies; \(\mathcal{H}\): dependency homophily; \(\langle k \rangle\): mean degree.\label{tab:datasets}}
\renewcommand{\arraystretch}{1.2}
\begin{adjustbox}{max width=\columnwidth}
\begin{tabular}{c l l r r r c r}
\toprule
& System & Version & $M$ & Files & $|E|$ & $\mathcal{H}$ & $\langle k \rangle$ \\
\midrule

\multirow{6}{*}{\rotatebox[origin=c]{90}{Java}}
& A.UML   & v0.24 & 14 & 766 & 3\,754 & 0.55 & 9.80 \\
& C.Img   & v1.0a2 & 21 & 329 & 1\,565 & 0.61 & 9.51 \\
& JabRef  & v3.7 & 6 & 1\,013 & 6\,126 & 0.56 & 12.09 \\
& Lucene  & r1075001 & 7 & 1\,054 & 6\,424 & 0.68 & 12.19 \\
& SH-3D   & r002382 & 9 & 167 & 1\,220 & 0.36 & 14.61 \\
& T.Mates & v5.11 & 15 & 778 & 6\,721 & 0.34 & 17.28 \\
\midrule

\multirow{2}{*}{\rotatebox[origin=c]{90}{C}}
& Bash & 4.2 & 14 & 292 & 1\,010 & 0.58 & 6.92 \\
& HDF  & 0e196f & 9 & 153 & 535 & 0.35 & 6.99 \\
\midrule

\multirow{2}{*}{\rotatebox[origin=c]{90}{C++}}
& HDC    & 46ff87 & 11 & 207 & 543 & 0.63 & 5.25 \\
& Chrome & svn-171054 & 69 & 18\,343 & 145\,492 & 0.58 & 15.86 \\

\bottomrule
\end{tabular}
\end{adjustbox}
\end{table}

The Java systems come from the benchmark of Olsson et al.~\cite{olsson2022automatically}. JabRef and TeamMates (T.Mates) were studied at the SAEroCon workshop in 2016 and 2017, where system experts provided the architectures and mappings\footnote{\label{fnsaerocon}\href{https://github.com/sebastianherold/SAEroConRepo}{https://github.com/sebastianherold/SAEroConRepo}}. Lucene was studied by Lenhard et al.~\cite{lenhard2019exploring}, and the Commons Imaging (C.Img) mapping was contributed as a test case for s4rdm3x~\cite{olsson2021s4rdm3x} \footnote{\label{fns4rdm3x} Java labels are available at: \href{https://github.com/h0bb3/s4rdm3x}{https://github.com/h0bb3/s4rdm3x}}.

For Bash, we use the mapping curated by Garcia et al.~\cite{garcia2013}, who recovered an initial decomposition from documentation and source code and refined it with system experts through a certification process. The same work provides the Chromium mapping, developed with developers over approximately two years~\cite{deps_impact}.

The HDF and HDC mappings were developed with industry partners and are part of the SARIF benchmark of Zhang et al.~\cite{sarif}, available in its replication package\footnote{\label{fnsarif}\href{https://github.com/anonymous2f4a9d/SARIF_FSE23/tree/main}{https://github.com/anonymous2f4a9d/SARIF\_FSE23}}.

\subsection{Experimental Configurations}

We repeat each experiment 100 times, each with a different initial labeled set. In each run, we randomly sample 5\% of the file nodes, with a minimum of 30 and a maximum of 200 files, and add one file from each GT module not represented in the sample. We use the same 100 initial labeled sets across all experiments and configurations, which allows paired comparisons. This procedure imposes a bounded labeling budget rather than a constant labeled proportion: before the module-coverage additions, the random sample covers approximately 9--20\% of the files in the five systems subject to the floor (C.Img, SH-3D, Bash, HDF, and HDC) and approximately 1\% in Chromium, the only system subject to the cap.

\parg{Input representations.}
For each system, Word2Vec embeddings are 128-dimensional vectors trained for 20 epochs with a context window of five on the file names and identifiers.

Node2Vec learns 128-dimensional vectors from uniform random walks ($p=q=1$) on the undirected file-level dependency graph, with walk length 30, 200 walks per file, window 15, five skip-gram epochs, and five negative samples. Chromium uses 100 walks per file and one negative sample.

For \ft{}, UniXcoder is fine-tuned separately for each system with an InfoNCE temperature of $0.05$, AdamW with a learning rate of $2\times10^{-5}$, a batch size of eight, and two epochs with a fixed seed. Only the upper four of the twelve encoder layers are updated, while the token embeddings and the lower eight layers stay frozen. Both UniXcoder variants encode file inputs with a 512-token limit and member inputs in overlapping 256-token windows with a stride of 128, mean-pooled before normalization, producing 768-dimensional features.

All input representations are generated once per system and reused in every run and experiment.

\parg{Training models.}
All models are trained with Adam at a learning rate of $0.001$ for 200 epochs. The MLP has two hidden layers of 256 dimensions, and GAT and HGT use one layer with 256 hidden dimensions and four attention heads.

\parg{Controlled comparisons.}
RQ1 trains the MLP on each of the four representations, which isolates the information in the file representation because the MLP uses no graph structure.

RQ2 trains HGT on all graph variants (Figure~\ref{fig:graph_variants_overview}) with Word2Vec, Node2Vec, and \ft{} features, using directed dependency edges. We additionally train all four HGT variants with two layers to test whether one layer limits what a file receives from its neighbors. We also train \textbf{FD}, an HGT model on graph F whose \ft{} file vector is extended with a binary folder-path vector, including ancestor folders. This vector uses the same tokenizer and vocabulary as the folder-node features (Section~\ref{sec:node_features}), with one dimension per distinct folder token in the system. FD provides folder information without folder nodes.

RQ3 consists of controlled comparisons on the file graph, with all four representations. For each predictor (MLP, GAT, and HGT), we compare settings with dependency neighbors disabled and enabled, and with different aggregation directions. For the neighbor MLP, enabling neighbors also increases the input width and parameter count. We keep the file representations, initial labeled sets, and training settings fixed and compare three settings. With neighbors \emph{disabled}, each file uses only its own representation. With \emph{directed} aggregation, a dependency edge from file $u$ to file $v$, where $u$ depends on $v$, carries information from $u$ to $v$. Each file therefore receives information from the files that depend on it. With \emph{undirected} aggregation, we also add the reverse edge, allowing each file to receive information from files it depends on. For the MLP, these settings determine which files contribute to the neighbor vector; for GAT and HGT, they determine which files send messages.

The MLP with neighbors disabled is the RQ1 classifier, and HGT with directed neighbors on the file graph is the F configuration of RQ2, so these runs are shared. In total, the experiments comprise 58,000 runs.

\subsubsection{Directory Hierarchy Baseline\label{sec:dir_predictor}}
To assess how much folder-based configurations exploit files in the same directory sharing a module, we use a rule-based predictor based on the directory hierarchy and the initial labeled files. For each unlabeled file, it assigns the most common module among labeled files in the same directory. If none are present, it searches the directory subtree and then the subtrees of its parent directories until it finds labeled files. Ties are resolved by the module's frequency in the complete initial labeled set and then alphabetically. If no labeled files are found, the predictor assigns the most common module in the complete initial labeled set. It uses the directory hierarchy, but not directory-name text, source code, embeddings, dependencies, or test labels.

We apply the predictor to all 100 initial labeled sets and compare it with HGT using \ft{} features on the directed FM and FFM graphs and with FD. For each system, we report the accuracy of the predictor, FM, FFM, and FD, which equals micro-F1 because each file has exactly one module label. For the files that the predictor assigns incorrectly, we also report the accuracy of each compared model on those files.

\subsection{Performance Evaluation}
The benchmark systems have imbalanced module sizes, so overall accuracy can hide poor performance on smaller modules. We therefore report both macro-F1 and micro-F1. F1 is the harmonic mean of precision and recall, which measure the correctness of predicted assignments and the coverage of true assignments, respectively. Macro-F1 averages the F1 scores of individual modules, giving each module equal weight regardless of size. Micro-F1 aggregates predictions across modules, giving each file equal weight. Reporting both captures performance across modules as well as overall file-level performance.

Because compared configurations use the same initial labeled sets, we pair runs by split. We use the two-sided Wilcoxon signed-rank test~\cite{wilcoxon}, a non-parametric test, to compare configurations separately for each system and metric without assuming normally distributed score differences. We apply Holm's correction across the ten systems for each metric and pair of compared configurations, using a significance level of $\alpha=0.05$.

To quantify the magnitude of the differences, we also compute Cliff's $\delta$~\cite{cliff} from the two sets of 100 scores. We classify its magnitude as negligible (N, $|\delta|<0.147$), small (S, $0.147\leq|\delta|<0.33$), medium (M, $0.33\leq|\delta|<0.474$), or large (L, $|\delta|\geq0.474$)~\cite{romano2006}.

\section{Results\label{sec:results}}
This section reports the results for code-to-architecture mapping across ten subject systems. RQ1 compares lexical, structural, and contextual file representations using an MLP. RQ2 examines how mapping performance changes when Word2Vec, Node2Vec, and \ft{} initialize HGT with file, member, and folder context, and analyzes errors relative to directory-based predictions. RQ3 uses controlled experiments to examine how using dependency neighbors changes mapping performance, and how the predictor and aggregation direction affect this change.

For RQ1 and RQ2, we report median macro- and micro-F1 scores over the 100 runs per system and configuration, and differences between these medians in F1 percentage points, with Cliff's $\delta$ alongside the significance test. For RQ3, we report the median of the paired differences between conditions over the 100 splits. Summary tables give the median of these per-system values across the ten systems. F1 is computed only on the held-out test files of each split.
\subsection{Comparing Lexical, Structural, and Contextual File Representations (RQ1)\label{sec:rq1}}

RQ1 examines the mapping performance of each file representation before additional graph-based message passing. Table~\ref{tab:rq1_mlp_features} reports the median macro-F1 and micro-F1 scores obtained using the MLP with Word2Vec, Node2Vec, and the two UniXcoder variants, \zs{} and \ft{}.

Table~\ref {tab:rq1_stats} summarizes the pairwise differences across all ten systems, with the number of systems favoring each method and the number of Holm-significant differences in each direction.

\begin{table}[htb]
\centering
\small
\renewcommand{\arraystretch}{1.1}
\caption{Median macro-F1/micro-F1 scores (\%) across 100 runs for the MLP using W2V, \zs{}, \ft{}, and N2V. Highlighted values indicate the highest score for each system and metric.\label{tab:rq1_mlp_features}}
\begin{adjustbox}{width=\columnwidth}
\setlength{\tabcolsep}{11pt}
\begin{tabular}{lcccc}
\toprule
System & W2V & \zs{} & \ft{} & N2V \\
\midrule
A.UML
& 34\,/\,57
& 36\,/\,60
& 46\,/\,69
& \best{54}\,/\,\best{72} \\

C.Img
& 23\,/\,39
& 33\,/\,52
& 42\,/\,62
& \best{71}\,/\,\best{80} \\

JabRef
& \best{44}\,/\,78
& 39\,/\,76
& 43\,/\,\best{79}
& 40\,/\,74 \\

Lucene
& 57\,/\,71
& 54\,/\,66
& 63\,/\,76
& \best{66}\,/\,\best{79} \\

SH-3D
& 22\,/\,41
& \best{48}\,/\,64
& 47\,/\,\best{67}
& 43\,/\,58 \\

T.Mates
& 30\,/\,55
& 37\,/\,68
& \best{51}\,/\,\best{79}
& 41\,/\,68 \\

Bash
& 7\,/\,29
& 23\,/\,42
& 27\,/\,49
& \best{42}\,/\,\best{61} \\

HDF
& 8\,/\,22
& 42\,/\,44
& 46\,/\,48
& \best{53}\,/\,\best{57} \\

Chrome
& 14\,/\,48
& 14\,/\,48
& 31\,/\,66
& \best{41}\,/\,\best{71} \\

HDC
& 33\,/\,53
& 30\,/\,51
& \best{46}\,/\,\best{70}
& 43\,/\,69 \\

\bottomrule
\end{tabular}
\end{adjustbox}
\end{table}

\begin{table}[h]
\centering
\caption{Summary of pairwise differences in F1 points across ten systems. $\Delta$ is the median difference (first method minus second). $+$ve is the number of systems favoring the first method, followed in parentheses by the numbers of significant positive ($\uparrow$) and negative ($\downarrow$) differences.\label{tab:rq1_stats}}
\begin{adjustbox}{max width=\linewidth}
\renewcommand{\arraystretch}{1.1}
\setlength{\tabcolsep}{8pt}
\begin{tabular}{lrcrc}
\toprule
& \multicolumn{2}{c}{Macro F1} & \multicolumn{2}{c}{Micro F1} \\
\cmidrule(lr){2-3}\cmidrule(lr){4-5}
Comparison & $\Delta$ & $+$ve ($\uparrow$/$\downarrow$) & $\Delta$ & $+$ve ($\uparrow$/$\downarrow$) \\
\midrule
\zs{}$-$W2V   & $+$5.2 & 6 (6/4) & $+$7.5 & 6 (6/3) \\
\ft{}$-$\zs{} & $+$9.1 & 9 (9/0) & $+$9.2 & 10 (10/0) \\
\ft{}$-$W2V   & $+$17.5 & 9 (9/0) & $+$19.2 & 10 (9/0) \\
\ft{}$-$N2V   & \best{$-$4.9} & 4 (4/6) & \best{$-$3.3} & 3 (3/6) \\
N2V$-$W2V     & $+$20.3 & 9 (9/1) & $+$16.2 & 9 (9/1) \\
\bottomrule
\end{tabular}
\end{adjustbox}
\end{table}

Table~\ref{tab:rq1_mlp_features} shows that macro-F1 is lower than micro-F1 for every system and representation. Node2Vec achieves the highest median macro- and micro-F1 scores on six systems. On the other four systems, \ft{} obtains the highest median micro-F1, while the highest median macro-F1 is held by \ft{} on T.Mates and HDC, Word2Vec on JabRef, and \zs{} on SH-3D.

We first compare Word2Vec and \zs{}, the two representations that do not encode dependencies. They do not separate consistently: the significant differences fall in both directions, although \zs{} is favored on more systems. On macro-F1, \zs{} is significantly stronger on A.UML (medium effect), C.Img, SH-3D, T.Mates, Bash, and HDF (large effects), whereas Word2Vec is significantly stronger on JabRef (large effect), Lucene and Chrome (medium effects), and HDC (small effect). On micro-F1, \zs{} is significantly stronger on the same six systems, and Word2Vec is significantly stronger on JabRef and Lucene (large effects) and HDC (small effect); the difference in Chrome is not significant.

Fine-tuning significantly improves UniXcoder over \zs{} on micro-F1 for all ten systems and on macro-F1 for nine, predominantly with large effects. The effects are medium on HDF (both metrics) and on SH-3D micro-F1; the macro-F1 difference on SH-3D is not significant.

Both representations encoding dependency information are generally stronger than the lexical representation. Node2Vec significantly outperforms Word2Vec on both metrics for nine systems, with JabRef as the exception, where Word2Vec is significantly stronger. Similarly, \ft{} significantly outperforms Word2Vec on both metrics for nine systems; on JabRef, the difference is not significant on either metric.

Node2Vec is significantly stronger than \ft{} on both metrics for A.UML, C.Img, Lucene, Bash, HDF, and Chrome, whereas \ft{} is significantly stronger on JabRef, SH-3D, and T.Mates; most of these differences have large effects. On HDC, \ft{} is significantly stronger on macro-F1 with a small effect, and the micro-F1 difference is not significant.

\begin{GrayBox}{Answer to RQ1.}
Representations that encode dependencies, either directly through random walks (Node2Vec) or through dependency-guided fine-tuning of UniXcoder, provide the strongest mapping performance overall. Fine-tuning consistently improves UniXcoder over its zero-shot representation and produces the best macro-F1 on two systems. Word2Vec and \zs{}, which do not encode dependencies, differ from each other in both directions, and Word2Vec is weaker than Node2Vec and \ft{} on most systems; Word2Vec nevertheless remains stronger than Node2Vec on JabRef. Thus, no single file representation performs best across all systems and metrics.
\end{GrayBox}
\subsection{Message Passing over Dependency Edges, Member Nodes, and Folder Nodes (RQ2) \label{sec:rq2}}

\begin{figure*}[htb]
\centering
\includegraphics[width=1\textwidth]{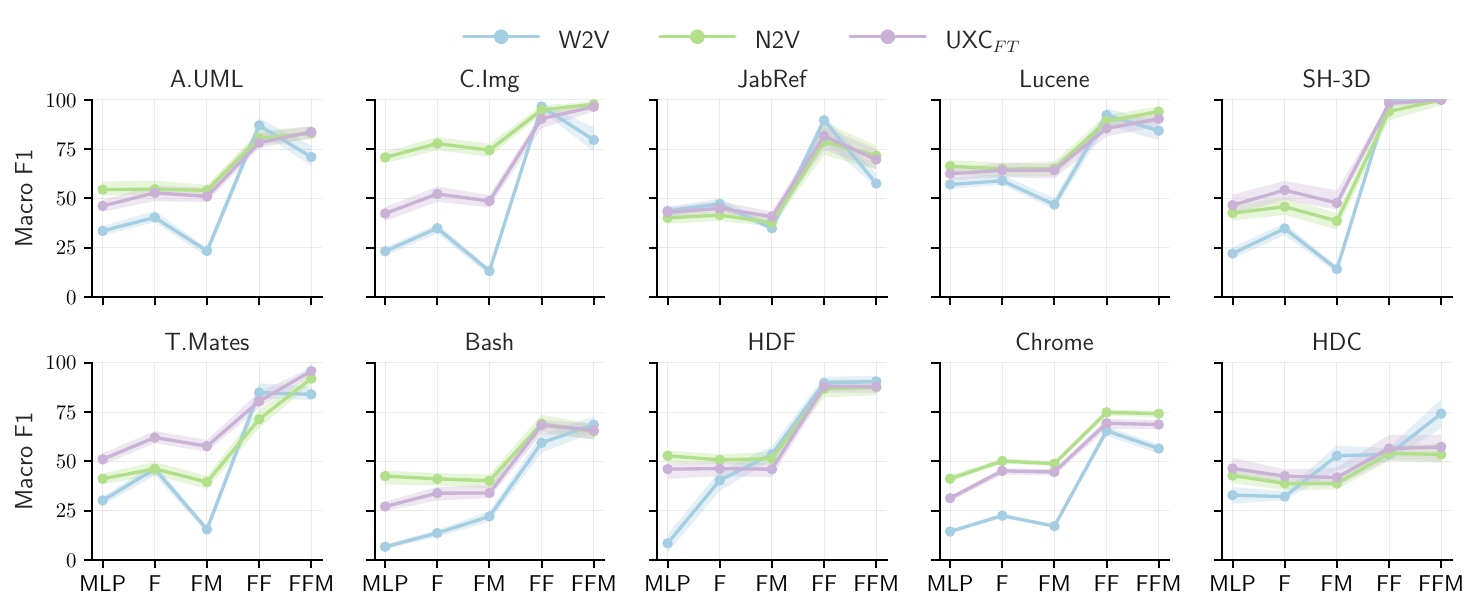}
\includegraphics[width=1\textwidth]{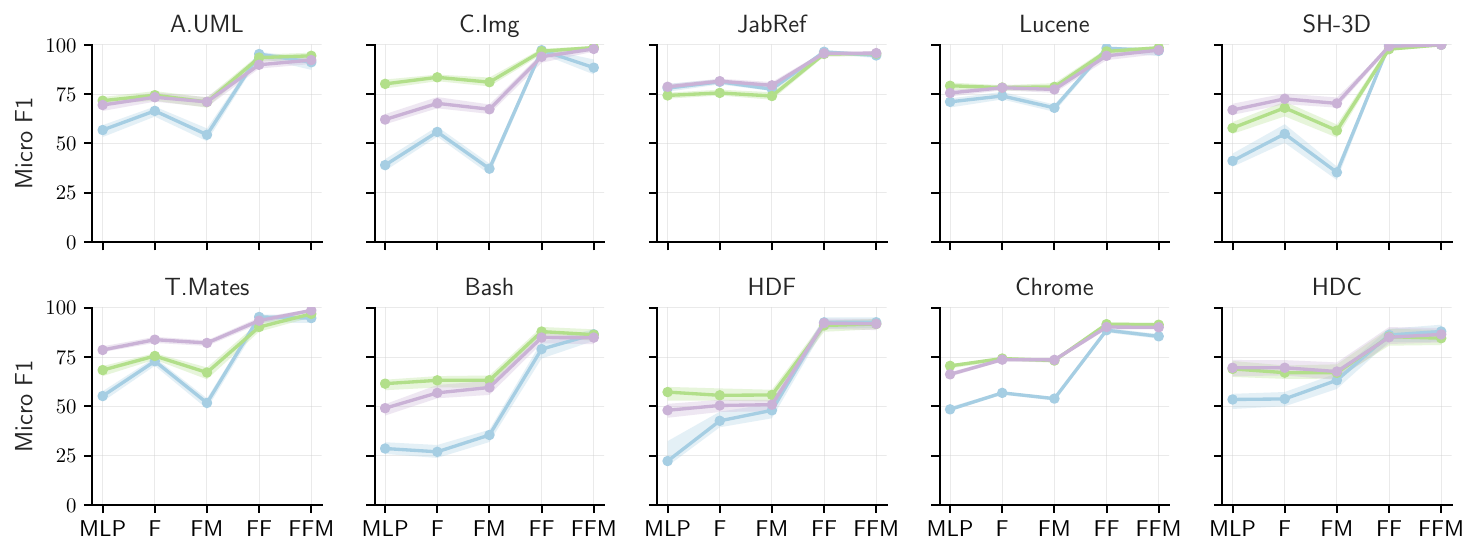}
\caption{Median macro-F1 (top) and micro-F1 (bottom) for each representation, per system, across the MLP and HGT with increasing graph structure (F, FF, FM, FFM). Shaded areas denote interquartile ranges across the 100 runs. \label{fig:rq2_lineplot}}
\end{figure*}

RQ2 examines how mapping performance changes for each representation when message passing over the file dependency graph is added and the graph is extended with folder and member nodes. Figure~\ref{fig:rq2_lineplot} reports median macro- and micro-F1 for the MLP and HGT under F, FF, FM, and FFM. We use Word2Vec, Node2Vec, and the fine-tuned variant \ft{}, which improves over \zs{} on nearly all systems in RQ1.

Table~\ref{tab:rq2_steps} summarizes five steps: MLP to F, adding members (F to FM, FF to FFM), and adding folders (F to FF, FM to FFM). For each system, $\Delta$ is the difference between the two configuration medians over 100 runs. The table reports the median of these ten system-level differences, the number of systems with a positive difference, and the number of systems with a significant gain or loss.
\begin{table*}[!t]
\centering
\caption{Change in F1 as graph structure is added, by representation. Each cell gives the median over ten systems of the difference between configuration medians in F1 points, followed in parentheses by the number of systems with a positive difference and the number of significant positive/negative differences. Blue shading indicates gains and purple indicates losses, with darker shades indicating larger magnitudes.\label{tab:rq2_steps}}
\begin{adjustbox}{max width=\textwidth}
\setlength{\tabcolsep}{10pt}
\renewcommand{\arraystretch}{1.2}
\begin{tabular}{lrrrrrr}
\toprule
& \multicolumn{2}{c}{W2V} & \multicolumn{2}{c}{N2V} & \multicolumn{2}{c}{\ft{}} \\
\cmidrule(lr){2-3}\cmidrule(lr){4-5}\cmidrule(lr){6-7}
Step & Macro F1 & Micro F1 & Macro F1 & Micro F1 & Macro F1 & Micro F1 \\
\midrule
F $-$ MLP & \gc{100}+7.5 (9; 9/0) & \gc{100}+9.0 (9; 8/0) & \gc{25}+0.8 (6; 5/3) & \gc{45}+2.2 (7; 7/2) & \gc{100}+6.7 (9; 8/1) & \gc{70}+4.6 (9; 9/0) \\
FM $-$ F & \rc{100}$-$12.2 (3; 3/7) & \rc{70}$-$4.9 (3; 3/7) & \rc{45}$-$1.2 (2; 0/5) & \rc{45}$-$1.3 (3; 0/6) & \rc{45}$-$1.2 (2; 1/6) & \rc{45}$-$1.8 (2; 1/7) \\
FFM $-$ FF & \rc{70}$-$4.4 (3; 2/5) & \rc{25}$-$1.0 (3; 2/5) & \gc{45}+1.6 (6; 5/3) & \gc{25}+0.8 (7; 5/1) & \gc{45}+1.3 (6; 5/3) & \gc{45}+1.1 (7; 5/0) \\
FF $-$ F & \gc{100}+44.4 (10; 10/0) & \gc{100}+32.1 (10; 10/0) & \gc{100}+25.6 (10; 10/0) & \gc{100}+18.7 (10; 10/0) & \gc{100}+29.9 (10; 10/0) & \gc{100}+16.5 (10; 10/0) \\
FFM $-$ FM & \gc{100}+43.0 (10; 10/0) & \gc{100}+39.9 (10; 10/0) & \gc{100}+28.9 (10; 10/0) & \gc{100}+22.3 (10; 10/0) & \gc{100}+32.3 (10; 10/0) & \gc{100}+20.6 (10; 10/0) \\
\bottomrule
\end{tabular}
\end{adjustbox}
\end{table*}

\parg{File-level HGT versus the MLP.} Moving from the MLP to file-level HGT improves Word2Vec and \ft{} more than Node2Vec. For Word2Vec and \ft{}, the improvement is positive on nine systems on both metrics, and the only significant loss is \ft{} on HDC macro-F1. For Node2Vec, the changes are small, and significant losses occur on Lucene and HDC for both metrics and on HDF for macro-F1, but significant gains still outnumber losses on both metrics.

\parg{Member-level context.} Adding member nodes to F lowers median macro-F1 and micro-F1 for every representation, most strongly for Word2Vec, with large effects on all ten systems. Significant macro-F1 losses outnumber gains for every representation. Adding member nodes also changes how features enter the model, depending on the representation (Section~\ref{sec:node_features}); for example, the \ft{} member vectors come from an encoder fine-tuned on file-level text. 

In addition, with one HGT layer, member-level dependencies cannot affect file predictions, so we test two layers in Section~\ref{sec:mp_depth}. These results therefore concern the evaluated member configuration, not member information in general.

\parg{Folder-level context.}
Adding folder nodes produces the largest and most consistent changes. FF improves over F, and FFM improves over FM, on both metrics for every system and representation. All 120 system--representation--metric comparisons are significant with a large effect, and each improves in at least 96 of the 100 paired runs.

Within the folder-containing configurations, the effect of adding member nodes depends on the representation. FFM improves over FF for Node2Vec and \ft{} on both metrics, with small median gains, but lowers both metrics for Word2Vec. The significant macro-F1 losses for Node2Vec and \ft{} occur on JabRef, Bash, and Chrome. For micro-F1, Node2Vec loses significantly only on Bash, and \ft{} on no system.

\subsubsection{Depth of message-passing layers \label{sec:mp_depth}}
To test whether using one HGT layer limits the results, we repeated F, FF, FM, and FFM with two layers, using the same labeled sets, representations, and training settings. Table~\ref{tab:rq2_depth} reports the results.

\begin{table}[h]
\centering
\caption{\fontsize{9.8}{12}\selectfont Effect of a second message-passing layer on HGT, by input and graph variant. $\Delta$ is the median across ten systems of the change in median F1 (points) from one to two layers; positive favors two layers. +ve is the number of systems with a positive change, followed in parentheses by the numbers of significant gains ($\uparrow$) and losses ($\downarrow$). Shading marks at least five significant gains and at least twice as many gains as losses.\label{tab:rq2_depth}}

\setlength{\tabcolsep}{8pt}
\begin{adjustbox}{max width=\columnwidth}
\begin{tabular}{llrcrc}
\toprule
& & \multicolumn{2}{c}{Macro-F1} & \multicolumn{2}{c}{Micro-F1} \\
\cmidrule(lr){3-4}\cmidrule(lr){5-6}
Input & Graph & $\Delta$ & +ve ($\uparrow$/$\downarrow$) & $\Delta$ & +ve ($\uparrow$/$\downarrow$) \\
\midrule
\multirow{4}{*}{W2V} & F & +0.5 & 6 (3/4) & 0.0 & 6 (4/3) \\
  & FF & \best{+2.4} & \best{8 (6/1)} & \best{+0.5} & \best{6 (6/0)} \\
  & FM & \best{+3.5} & \best{6 (6/3)} & \best{+2.2} & \best{8 (7/2)} \\
  & FFM & \best{+10.3} & \best{6 (6/3)} & \best{+1.1} & \best{6 (6/3)} \\
\midrule
\multirow{4}{*}{N2V} & F & $-$0.6 & 4 (2/5) & $-$0.1 & 4 (2/4) \\
  & FF & \best{+1.8} & \best{7 (5/2)} & +0.4 & 6 (4/3) \\
  & FM & 0.0 & 5 (3/3) & +0.1 & 5 (3/3) \\
  & FFM & +0.3 & 5 (4/3) & $-$0.2 & 3 (2/4) \\
\midrule
\multirow{4}{*}{\ft{}} & F & 0.0 & 6 (2/1) & \best{+0.8} & \best{8 (7/1)} \\
  & FF & \best{+3.4} & \best{8 (8/2)} & \best{+1.4} & \best{10 (7/0)} \\
  & FM & $-$0.2 & 5 (1/2) & 0.0 & 5 (3/3) \\
  & FFM & \best{+2.6} & \best{7 (7/1)} & \best{+1.0} & \best{8 (8/1)} \\
\bottomrule
\end{tabular}
\end{adjustbox}
\end{table}

For Word2Vec, the largest median macro-F1 improvements occur in FM and FFM. For \ft{}, FF improves more than FFM, so the improvement is not specific to member nodes. Node2Vec shows little change in either member-containing configuration, but F is significantly lower on macro-F1 on five systems.

At matched two-layer depth, adding members to F still lowers median macro-F1 for all three representations. Adding members to FF remains negative for Word2Vec, while the small median gain for \ft{} shrinks and that for Node2Vec becomes slightly negative. Folder nodes retain the largest gains: both folder additions improve both metrics on all ten systems for every representation, with significant large effects. Thus, additional depth improves some configurations but does not establish a consistent benefit from adding member nodes.

We also examined whether the macro-F1 change from one to two message-passing layers in FM and FFM relates to average file-level degree across the ten systems. Exploratory Kendall's $\tau$ analyses~\cite{kendall} show positive associations in FFM across all three representations, but in FM mainly with Word2Vec.

\subsubsection{Organizational information and the directory baseline\label{sec:directory_errors}}

The large gains from adding folder nodes show that directory organization is informative for the evaluated mappings. We first test whether these gains require folder nodes using FD, which adds folder-path features directly to the \ft{} file vectors in graph F. We then examine whether the gains also help files misclassified by directory information alone, comparing FM, FFM, and FD against the directory hierarchy baseline (Section~\ref{sec:dir_predictor}). FM and FFM differ only in the inclusion of folder nodes. Table~\ref{tab:dir_baseline} reports overall accuracy for the baseline and the three models, and model accuracy on files misclassified by the baseline.
\begin{table}[h]
\centering
\caption{\fontsize{9.4}{12}\selectfont Accuracy of the directory baseline (Dir.) and of HGT with \ft{} features on FM, FFM, and FD, median over 100 runs. $n$ is the median number of test files per split that the baseline misclassifies; the columns under it give model accuracy on those files, over splits with at least one baseline error. Purity is the share of files that belong to the majority module of their directory.\label{tab:dir_baseline}}
\begin{adjustbox}{max width=\linewidth}
\normalsize
\renewcommand{\arraystretch}{1.1}
\begin{tabular}{lrrrrrrrrr}
\toprule
& \multicolumn{4}{c}{Accuracy (\%)} & \multicolumn{4}{c}{Baseline misclassifications} & \\
\cmidrule(lr){2-5}\cmidrule(lr){6-9}
System & Dir. & FM & FFM & FD & $n$ & FM & FFM & FD & Purity \\
\midrule
A.UML   & 99  & 71 & 92  & 93 & 11   & 9.1  & 21.7 & 22.5 & 100.0 \\
C.Img   & 100 & 67 & 98  & 99 & 0    & --   & --   & --   & 100.0 \\
JabRef  & 97  & 79 & 96  & 93 & 27   & 45.3 & 26.3 & 25.0 & 99.1  \\
Lucene  & 92  & 77 & 97  & 95 & 86 & 52.9 & 86.9 & 81.7 & 100.0 \\
SH-3D   & 100 & 70 & 100 & 99 & 0    & --   & --   & --   & 100.0 \\
T.Mates & 100 & 82 & 99  & 99 & 0    & --   & --   & --   & 100.0 \\
Bash    & 92  & 60 & 85  & 86 & 21   & 24.6 & 26.1 & 27.8 & 91.8  \\
HDF     & 93  & 51 & 92  & 93 & 8    & 27.3 & 50.0 & 40.0 & 100.0 \\
Chrome  & 97  & 74 & 90  & 94 & 574  & 19.8 & 20.0 & 23.3 & 99.9  \\
HDC     & 89  & 68 & 87  & 87 & 19   & 20.0 & 12.5 & 10.8 & 96.1  \\
\bottomrule
\end{tabular}
\end{adjustbox}
\end{table}
FD improves over F on every system with a large effect. Compared with FD, FF has lower median macro-F1 on eight systems, significantly so on six, and no significant gains. Folder nodes and relations therefore do not improve on supplying folder-path information directly as file features in this configuration.

The baseline reaches 89 to 100\% accuracy on every system. Without folder nodes, FM is 14 to 43 points below it. FFM and FD are within 7 points of it on nine systems and exceed it only on Lucene. Folder information therefore brings both FFM and FD close to the accuracy of the directory hierarchy baseline on most systems.

Directory membership closely matches module membership: on eight systems, 99 to 100\% of files belong to the majority module of their directory, and only Bash and HDC have lower purity. Files that depart from the majority module of their directory are therefore rare on most systems, which limits what the analysis can show about them.

On the files the baseline gets wrong, FFM gets most of them wrong on A.UML, JabRef, Bash, Chrome, and HDC, about half on HDF, and most of them right only on Lucene. Compared with FM, FFM is more accurate on these files on A.UML, HDF, and Lucene, similar on Bash and Chrome, and less accurate on JabRef and HDC. JabRef illustrates the distinction between overall accuracy and accuracy on baseline errors: FFM improves over FM by 16 percentage points overall but falls by 19 points on these files. FD is within 10 points of FFM on these files on every system. This analysis does not isolate files whose module differs from the majority module of their directory, so it does not show how well FFM handles such exceptions.

\begin{GrayBox}{Answer to RQ2.}
The effect on mapping performance depends on what is added and on the representation. Moving from the MLP to file-level HGT improves Word2Vec and \ft{} more than Node2Vec. Member nodes give no consistent gain and often lower performance, most strongly for Word2Vec. Folder nodes provide the largest gains, with significant large effects on all ten systems for all three representations. The findings for member and folder nodes remain with two HGT layers.

The directory hierarchy baseline achieves accuracy similar to FFM on most systems, and folder nodes do not consistently outperform folder information supplied directly as file features (FD). On most systems with directory-baseline errors, FFM misclassifies most of these files, and on some it is less accurate than FM. Strong folder-based scores therefore do not establish reliable performance on directory-baseline errors, and the analysis does not isolate files whose directory location is insufficient or misleading.
\end{GrayBox}
\subsection{Dependency Neighbors, Predictors, and Aggregation Direction (RQ3)\label{sec:rq3}}

RQ2 shows that the effect of graph context depends on the representation, configuration, and system. However, its comparison of the MLP with file-level HGT changes both the predictor and the availability of dependency messages. RQ3 therefore compares each predictor with dependency neighbors disabled and enabled, using HGT, GAT, and a neighbor MLP (\mlp{}). It also compares directed and undirected aggregation to examine whether the benefit depends on how dependency neighbors are aggregated. We evaluate all three predictors with all four input representations.

Differences are paired changes in F1 points, enabled minus disabled; positive values favor neighbor information. For direction, differences are undirected minus directed; positive values favor undirected neighbors. We report the median paired difference over the 100 splits. With neighbors disabled, the neighbor MLP reduces to the MLP of RQ1, and GAT and HGT receive no dependency messages.

\parg{How does the benefit of dependency neighbors vary across predictors and representations?} Table~\ref{tab:rq3_reps} shows the change in F1 when neighbor information is enabled, for each predictor and input representation. With undirected neighbors, the median change is positive in every cell on both metrics. GAT and the \mlp{} improve on at least eight systems in every cell and are never significantly lower, whereas HGT gains least with every representation and is significantly lower on a few systems with Node2Vec and Word2Vec. With directed neighbors, the median macro-F1 is lower with Node2Vec for all three predictors, by 0.6 to 2.8 points, and significant losses occur on up to six systems. For GAT and the \mlp{}, the gain from undirected neighbors is smallest with Node2Vec and largest with \zs{}.

This comparison differs from RQ2, where file-level message passing (F) is compared with the MLP, which changes the predictor together with the available structure. Here the predictor is fixed, and dependency neighbors are enabled and disabled, so a gain of F over the MLP does not imply that HGT benefits from dependency neighbors. With Node2Vec, for example, F raises the median macro-F1 over the MLP, whereas HGT with directed neighbors is lower than HGT without neighbors.

\begin{table*}[t]
\centering
\caption{Changes in F1 from enabling neighbor information (D$-$off, UD$-$off) and switching from directed to undirected aggregation (UD$-$D), by model and input representation. Each cell reports the median across ten systems of the per-system median paired change over runs, in F1 points. Parentheses follow the format (systems with a positive median change; significant gains/significant losses). D: directed (source-to-target); UD: undirected; off: neighbor information disabled. Blue shading indicates gains and purple indicates losses, with darker shades indicating larger magnitudes.\label{tab:rq3_reps}}

\begin{adjustbox}{max width=\textwidth}
\renewcommand{\arraystretch}{1.1}
\normalsize
\begin{tabular}{llrrrrrrrr}
\toprule
& & \multicolumn{2}{c}{N2V} & \multicolumn{2}{c}{W2V} & \multicolumn{2}{c}{\zs{}} & \multicolumn{2}{c}{\ft{}} \\
\cmidrule(lr){3-4}\cmidrule(lr){5-6}\cmidrule(lr){7-8}\cmidrule(lr){9-10}
Model & Contrast & Macro F1 & Micro F1 & Macro F1 & Micro F1 & Macro F1 & Micro F1 & Macro F1 & Micro F1 \\
\midrule
\multirow{3}{*}{HGT} & D$-$off & \rc{25}$-$0.6 (4; 4/6) & \gc{45}+1.6 (6; 6/4) & \gc{25}+0.2 (5; 4/5) & \gc{45}+1.8 (7; 7/3) & \gc{45}+1.6 (9; 7/0) & \gc{45}+2.7 (10; 10/0) & \gc{25}+0.7 (7; 6/3) & \gc{45}+1.8 (8; 8/0) \\
 & UD$-$off & \gc{25}+0.5 (8; 5/2) & \gc{25}+0.8 (9; 8/0) & \gc{45}+2.9 (7; 7/3) & \gc{70}+4.5 (8; 8/2) & \gc{45}+1.8 (10; 10/0) & \gc{45}+2.5 (10; 10/0) & \gc{25}+0.9 (10; 8/0) & \gc{45}+1.6 (10; 10/0) \\
 & UD$-$D & \gc{25}+0.3 (6; 5/4) & \gc{25}+0.1 (5; 3/4) & \gc{25}+0.7 (7; 4/3) & \gc{45}+1.5 (6; 5/4) & \rc{25}0.0 (5; 4/3) & \gc{25}+0.1 (5; 5/4) & \gc{25}+0.6 (6; 5/3) & \gc{25}+0.2 (5; 4/4) \\
\addlinespace[3pt]

\multirow{3}{*}{GAT} & D$-$off & \rc{45}$-$2.8 (4; 3/6) & \rc{25}$-$0.8 (3; 2/5) & \gc{45}+1.5 (8; 6/2) & \gc{45}+2.4 (7; 7/1) & \gc{70}+3.8 (10; 9/0) & \gc{100}+6.2 (10; 10/0) & \gc{25}+0.7 (6; 4/2) & \gc{45}+2.2 (8; 7/1) \\
 & UD$-$off & \gc{45}+1.4 (9; 8/0) & \gc{45}+1.4 (10; 10/0) & \gc{70}+5.6 (8; 7/0) & \gc{100}+6.7 (8; 8/0) & \gc{100}+8.2 (10; 10/0) & \gc{100}+9.5 (10; 10/0) & \gc{70}+3.8 (10; 8/0) & \gc{70}+4.1 (10; 10/0) \\
 & UD$-$D & \gc{70}+3.8 (7; 6/2) & \gc{45}+2.2 (8; 8/1) & \gc{45}+1.9 (7; 5/3) & \gc{45}+1.6 (7; 5/3) & \gc{70}+3.3 (6; 6/4) & \gc{45}+2.2 (6; 6/2) & \gc{45}+3.0 (6; 6/4) & \gc{45}+1.3 (7; 6/1) \\
\addlinespace[3pt]

\multirow{3}{*}{\mlp{}} & D$-$off & \rc{25}$-$0.6 (4; 4/5) & \gc{25}+0.8 (6; 5/4) & \gc{70}+3.4 (9; 9/1) & \gc{45}+2.7 (9; 8/1) & \gc{70}+4.8 (8; 7/1) & \gc{70}+5.1 (9; 9/0) & \gc{45}+2.8 (7; 6/2) & \gc{45}+2.9 (7; 7/2) \\
 & UD$-$off & \gc{45}+1.2 (10; 8/0) & \gc{45}+1.2 (9; 8/0) & \gc{70}+5.1 (10; 10/0) & \gc{70}+4.9 (10; 10/0) & \gc{100}+7.7 (10; 10/0) & \gc{100}+7.3 (10; 10/0) & \gc{70}+5.8 (10; 10/0) & \gc{70}+4.6 (10; 10/0) \\
 & UD$-$D & \gc{45}+1.6 (6; 6/4) & \gc{25}+0.4 (6; 5/3) & \gc{45}+1.3 (6; 6/3) & \gc{45}+1.9 (6; 6/2) & \gc{70}+3.5 (6; 6/4) & \gc{45}+2.5 (6; 6/4) & \gc{70}+3.1 (6; 6/3) & \gc{45}+1.3 (7; 7/3) \\
\bottomrule
\end{tabular}
\end{adjustbox}
\end{table*}

\begin{figure}[h]
\centering
\includegraphics[width=1\columnwidth]{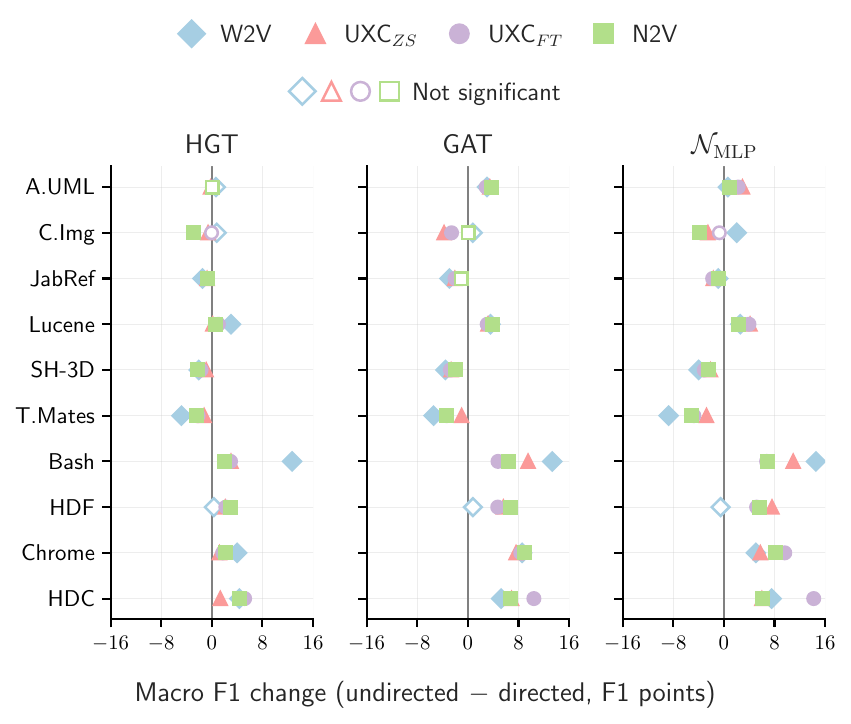}
\caption{Macro-F1 differences (undirected $-$ directed), in F1 points, by predictor and input representation. Each point shows median paired differences over 100 splits; negative values favor directed aggregation. Hollow markers indicate non-significant differences.\label{fig:rq3_direction}}
\end{figure}

\parg{Does the direction of neighbor aggregation matter?} Undirected aggregation gives the higher median in almost every cell of Table~\ref{tab:rq3_reps}, but the preferred direction depends on the system. Figure~\ref{fig:rq3_direction} shows gains from undirected aggregation on five to eight systems per combination of predictor and input representation, with significant differences in both directions.

On macro-F1, SH-3D, T.Mates, and JabRef score higher with directed aggregation in all 12 combinations, and C.Img in eight. Lucene, Bash, Chrome, and HDC score higher with undirected aggregation in all 12 combinations, and A.UML and HDF in 11. The largest differences are 14.6 points in favor of undirected aggregation (Bash, Word2Vec, \mlp{}) and 8.7 points in favor of directed aggregation (T.Mates, Word2Vec, \mlp{}). Micro-F1 shows the same directional pattern for seven systems, while JabRef, C.Img, and A.UML show mixed results. Adding reverse edges also expands the neighborhood, so this comparison does not isolate propagation direction from neighborhood size.

\begin{figure*}[h]
\centering
\includegraphics[width=0.98\textwidth]{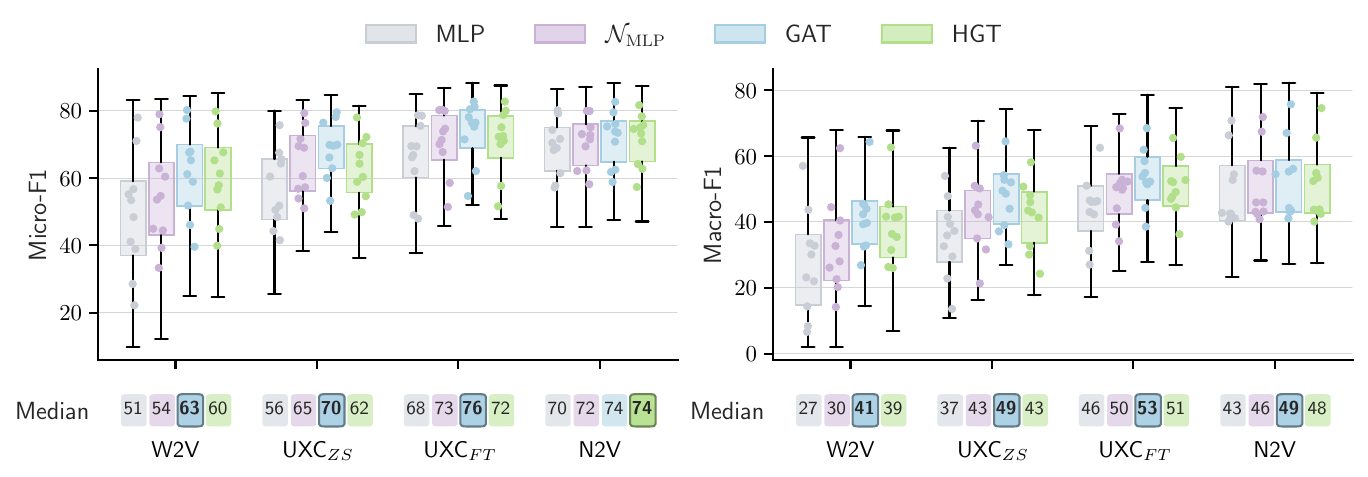}
\caption{Micro-F1 (left) and macro-F1 (right) of the MLP, \mlp{}, GAT, and HGT with undirected neighbors, by input representation. Each point is the median of one system over 100 runs; boxes show the spread across systems, and the values below are the median of medians across systems, with the highest in each group in bold.\label{fig:rq3_agg}}
\end{figure*}

\parg{Which predictor scores highest?} Figure~\ref{fig:rq3_agg} pools the ten systems with undirected neighbors: each point is the median of one system and the boxes show the spread across systems. GAT has the highest median macro-F1 with every representation and the highest median micro-F1 with every representation except Node2Vec, where HGT is marginally higher. GAT also scores highest most often: in each split we count which of the four predictors scores highest, and GAT has the largest share of wins, averaged over systems, with every representation and on both metrics. All three predictors with neighborhood aggregation exceed the MLP without neighbors in every median comparison. As without neighbors, \ft{} has the highest median macro-F1 across systems for every predictor, with GAT and \ft{} giving the highest scores overall.

Most of the variation in F1 lies between systems and not between runs of one system. The choice of predictor matters most for Word2Vec and \zs{}, the two representations without dependency information, and less for Node2Vec and \ft{}.

\begin{GrayBox}{Answer to RQ3.}
Dependency neighbors do not benefit all predictors equally. With undirected aggregation, median changes are positive for every predictor and representation, but GAT and the \mlp{} improve more consistently across systems than HGT, which shows significant losses in some cases. Directed aggregation is less reliable, particularly with Node2Vec, where median macro-F1 decreases for all three predictors. Although undirected aggregation generally gives larger gains, it is not preferable on every system: several systems consistently favor directed aggregation across predictors and representations. These comparisons also change neighborhood size, so they do not isolate direction alone. With undirected neighbors, GAT has the highest median macro-F1 across systems for every representation, and \ft{} remains the strongest representation by this measure for every predictor.
\end{GrayBox}

\section{Discussion\label{sec:discussion}}
The research questions examine where architectural signal resides and how a mapping model can use it. RQ1 compares file representations before message passing, RQ2 examines file, member, and folder context and compares the folder gains with a directory-based rule, and RQ3 examines how predictor choice and aggregation direction affect the contribution of dependency neighbors. This section discusses these findings together.

\subsubsection*{Dependency Information in File Representations}
Dependency information can support mapping before message passing, through the graph position encoded by Node2Vec or through dependency-guided adaptation of a code encoder. The advantage of these representations over Word2Vec, and of fine-tuned UniXcoder over its zero-shot counterpart, suggests that dependency information is useful beyond what the evaluated lexical and pretrained code representations provide. The representations also differ in training, vector size, and source content, so these comparisons are between representation pipelines and not between isolated information sources.

The usefulness of this signal varies with the relation between dependency structure and the reference architecture. Node2Vec’s micro-F1 is lower on the three least homophilic systems, SH-3D, HDF, and T.Mates (Table~\ref{tab:datasets}), with a median of $58$ compared with $72$ on the remaining seven. Node2Vec learns dependency proximity from files that co-occur on random walks, without using the GT mapping. When dependencies frequently cross module boundaries, files from different modules may co-occur on walks and receive similar vectors, making the modules harder to distinguish.

The advantage of fine-tuned UniXcoder over Node2Vec on SH-3D and T.Mates suggests that code content may provide complementary evidence in these mixed neighborhoods. HDF does not follow this pattern, so low homophily alone does not explain their relative performance; HDF is also the smallest system, and every representation has comparatively low micro-F1 on it. Homophily is used here to interpret the observed pattern, not to select a representation, since computing it requires the reference mapping.

Encoding dependencies in the initial representation and aggregating dependency neighbors during mapping are two uses of the same information source, and the second does not necessarily replace the first. With undirected aggregation, the median gains are generally larger for Word2Vec and zero-shot UniXcoder than for Node2Vec and fine-tuned UniXcoder, yet fine-tuned UniXcoder still has the highest median macro-F1 for every predictor. Conversely, useful dependency information in a representation does not guarantee a benefit from further aggregation: RQ3 shows that this benefit depends on both aggregation direction and predictor choice.

\subsubsection*{Folder Organization and Architectural Exceptions}
The gains from adding folder nodes in RQ2 show that directory organization helps reproduce the reference mapping. They do not show that this information must reach a file through message passing over folder nodes, since supplying the same information directly in the file vector matches or exceeds the accuracy of folder nodes on most systems. Extending a directory assignment is also different from identifying files that depart from it. Comparing FM, FFM, and FD on the directory baseline's errors helps assess this distinction (Section~\ref{sec:directory_errors}), but does not isolate architectural exceptions from errors caused by missing local supervision.

Lucene and HDF have pure directories under the reference mapping, so their baseline errors do not reflect conflicts between directory and module membership. They may instead arise from directories without labeled examples, although the table does not establish this. FFM recovers most baseline errors on Lucene and half on HDF, but only about a fifth on A.UML, whose directories are also pure. Directory purity therefore does not ensure that folder nodes recover baseline errors.

Bash and HDC have mixed directories. On baseline errors, FFM is similar to FM on Bash and less accurate on HDC. JabRef further illustrates why overall gains are insufficient: adding folder context raises overall accuracy while lowering accuracy on baseline errors.

The directory rule serves as a reference point for interpreting the learned models. Where directory organization closely follows the architecture, it already reproduces most of the mapping. On most evaluated systems, the added modeling complexity therefore brings little accuracy over the rule. Evaluating the additional value of a learned mapper requires examining both files in directories without labeled examples and files whose module differs from their directory's majority module.

\subsubsection*{Dependency Neighbors and Aggregation Direction}

The benefit of dependency neighbors depends on how the predictor aggregates them. With undirected aggregation, HGT gains least, while GAT achieves the highest median macro-F1 with every representation, although differences between predictors are small relative to variation across systems. GAT and HGT differ in their attention formulations and parameterizations, so the comparison does not isolate why GAT performs better. These findings concern the evaluated file-level predictors, not the usefulness of dependency types in general. RQ2 uses HGT for the folder- and member-aware configurations; whether the folder gains persist with other predictors remains untested.

The preferred aggregation direction appears to depend more on the system than on the predictor or representation, although the pattern is not uniform across metrics. Undirected aggregation generally gives larger gains, but several systems consistently favor directed aggregation. Agreement across predictors and representations describes the stability of these system-level preferences rather than providing independent evidence, since the comparisons share systems and splits.

Undirected aggregation adds reverse edges, so the comparison does not isolate propagation direction from neighborhood size. A reversed-edges-only setting would help distinguish these effects and remains for future work.

With Node2Vec, directed aggregation lowers median macro-F1 for all three predictors. Since its vectors were learned on an undirected graph, this may reflect a mismatch between the neighborhoods encoded during initialization and those used during prediction, but this explanation remains untested.

\subsubsection*{Connectivity and Neighbor Information}
Connectivity describes how much neighboring context is available, measured here by average node degree, while homophily describes whether linked files share a module. Few neighbors limit the context available to shallow message passing, whereas Node2Vec can encode more distant context through random walks. Limited connectivity may therefore constrain the benefit of additional aggregation even when the initial structural representation remains useful.

With Node2Vec and undirected aggregation, no predictor changes significantly on Bash and HDC, the two lowest-degree systems. The depth analysis is also consistent with this interpretation: higher average degree is associated with larger gains from a second message-passing layer, particularly in FFM.

Connectivity may also help explain the effect of aggregation direction. T.Mates and SH-3D combine low homophily with high average degree and favor directed aggregation in RQ3. Bash and HDC favor undirected aggregation, as does HDF, which has low homophily but much lower average degree than T.Mates and SH-3D. Reverse edges may supply missing context in sparse neighborhoods but introduce additional cross-module context in dense, low-homophily neighborhoods. This mechanism remains untested: the ten-system comparison cannot separate connectivity from homophily, system size, or label scarcity.

\subsubsection*{Implications for Evaluation}
Evaluation should establish what the input representation already provides, then isolate access to neighborhood information within a fixed predictor. Comparing a GNN model with an MLP changes both the predictor and the use of dependency-neighborhood aggregation. Enabling and disabling neighborhood aggregation within the same predictor family, while keeping the file representations fixed, provides a more targeted comparison. For the neighbor MLP, this comparison includes an increase in input width and parameter count. The same applies to folder information: a comparison of graph variants should include a model that receives that information directly as file features.

Micro-F1 and macro-F1 should be read together, since correctly mapping more files does not necessarily improve performance across modules. For tools intended to identify architectural exceptions, aggregate scores should also be accompanied by performance on files whose module differs from the majority module of their directory.

The intended use also matters. If mapping supports code reorganization, the existing directory structure may be something to reconsider rather than reproduce. Strong performance based on that structure does not establish which information would remain useful when the organization changes. The present study does not test this setting, but motivates evaluating information sources against the intended task rather than treating any representation or predictor as generally preferable.

\section{Related Work\label{sec:work}}

Architecture recovery and code-to-architecture mapping both infer a system's modular structure from implementation artifacts, such as static dependencies, code semantics, and directory organization. Recovery produces a decomposition without a reference architecture, while mapping assigns entities to the modules of a given one. We evaluate mapping and draw on recovery research for representations of dependencies, lexical content, and project organization.

Recovery techniques have historically been built around one dominant signal: dependency graphs, partitioned by search-based optimization~\cite{bunch,bunch_sahc_nahc,dagc,wca,mca_eca} or clustering~\cite{cct,hcc_2019,limbo,fca,neighborhood_tree_2019,gma_2021}, identifiers and comments~\cite{arc,zbr,risi2012fold,kargar2017semantic}, or file names and paths~\cite{file_clustering}. Each signal is noisy in isolation, since shared utilities introduce cross-module edges and reused terminology mixes entities from many modules, so hybrid methods combine sources~\cite{sc_misra,escrc,nsga3,sadat2019multi,ensemble_recovery,evol,sade,sarif}.

Mapping began with orphan adoption~\cite{tzerpos1997orphan} and the attraction functions of HuGMe~\cite{christl2005equipping,christl2007automated}, and later varied the signal and the learner: textual similarity~\cite{bittencourt_IR}, supervised classifiers~\cite{bibi2016supervised}, and semi-supervised Naive Bayes combining lexical and dependency-derived information~\cite{olsson2019semi,olsson2022automatically}. Text-based approaches compare classifiers for module assignment~\cite{florean2021textclassifierscomparison}, recommend mappings from natural-language architecture descriptions~\cite{InMap}, or integrate package- and class-level granularity~\cite{sinkala2023integrated}. More recent work uses TF-IDF, GloVe, and LLM representations~\cite{johansson2024mapping}, extracts architectural concepts from documentation and code~\cite{fuchss2025enabling}, or prompts LLMs to map implementation units to a reference architecture, which faces context-size limits and repeated processing costs on large codebases~\cite{rukmono_deductive}. Johansson and Caporuscio map functions and files to textual module descriptions with pre-trained embeddings and find files classified more reliably than functions, with performance associated with vocabulary overlap and class separability~\cite{johansson2026mapping}.

Representation-learning work on recovery encoded one or two sources at a time. NEGAR clusters Node2Vec embeddings of the dependency graph alone~\cite{negar}. Jabeen et al.~\cite{structsar} study dependencies alone and with folder organization, comparing random walk-based methods and variational graph autoencoders. GAER~\cite{gaer}, a graph autoencoder, and ARL~\cite{arl_2026}, a contrastively trained GAT, each fix folder and identifier tokens into one encoding without comparing alternatives. For mapping, GNNs were evaluated on homogeneous and dependency-type-aware graphs without folder or member nodes and without separating graph contribution from feature contribution~\cite{jabeen2025graph}. Its extension combines Word2Vec identifier embeddings and binary bag-of-words representations of folder path tokens with dependencies in one model~\cite{c2a_mapping}.

These approaches motivate our distinction between representing an information source and using it during prediction. Text-based mapping reports weaker results for functions than for files~\cite{johansson2026mapping}, but does not examine whether member information helps as context for the file. Our comparisons span lexical, structural, and contextual inputs but retain differences in their training procedures and feature content, so the matched dependency controls and directory-error analyses assess specific design choices and do not turn the broader comparison into an isolated ranking of information sources.
\section{Threats to Validity\label{sec:threats}}

\parg{Generalizability:} We evaluate ten open-source systems in Java, C, and C++, each with one reference architecture. The findings may not extend to other systems or alternative architectural decompositions. The study is transductive: representation learning and graph processing can use unlabeled test nodes and their edges. It therefore does not establish performance on newly arriving files outside the observed graph or on maintaining a mapping as the system evolves. Other information sources, including change history and dynamic dependencies, are outside the evaluated scope.

\parg{Dependencies and reference mappings:} Dependencies are extracted using static analysis, which may miss or misresolve dependencies. Preprocessing and build configuration can also affect the dependencies available for analysis. Because predictors that use neighbors depend on the extracted graph, such errors can affect their predictions. Models with neighbors enabled and disabled are trained separately. Their comparison therefore measures the effect of training with access to neighbor information, rather than removing neighbors from a fixed trained model.

We made a best-effort attempt to identify and obtain the source versions corresponding to the published mappings, using the available version information and documentation. Despite these efforts, some mismatches between the analyzed file sets and the mappings remain. These may reflect incomplete documentation of the original source versions or build configurations, which limits our ability to establish an exact correspondence between the analyzed source code and the published mappings.

Files can plausibly belong to more than one module, and annotators may assign them differently. Because the initial labeled set is sampled from the same mapping used for evaluation, an incorrect assignment can both mislead the model during training and penalize an architecturally plausible prediction.

\parg{Experimental variation:} We use 100 runs per configuration. These runs reuse the same systems and can include the same files, so they are not independent samples. With this many runs, small differences can reach statistical significance, so we report results per system and include effect sizes to distinguish the magnitude of a difference from its significance. The comparisons were not fixed in advance, so the analyses are exploratory and may favor patterns specific to this benchmark.

The runs use different initial labeled sets, but the contribution of stochastic model training to the observed variation was not assessed separately. Input representations were generated once per system and reused across runs, so variation arising from their generation was not assessed. The seeds governing model initialization and training were not retained, limiting exact reproducibility. Cliff's $\delta$ treats the paired scores as two independent groups, so it describes the separation of the two score distributions and not the paired differences.

\parg{Interpretation of results:} Performance measures agreement with the reference architecture, not whether a prediction is the only architecturally reasonable assignment. Folder organization may reflect the same decisions used to define that architecture, so strong directory-based predictions do not by themselves demonstrate recovery of architectural intent. The directory-error analysis applies only where the directory baseline makes errors, which excludes C.Img, SH-3D, and T.Mates in most or all splits, and it does not isolate files whose module differs from the majority module of their directory.
Message-passing depth was varied only for HGT in RQ2, with one and two layers, so the results do not show how deeper models or other predictors respond. It reports accuracy only, so it does not show how performance differs between small and large modules.

Relationships between performance and structural properties are reported as associations, not causal effects, since the benchmark does not allow these properties to be separated from one another or from system size and label availability. The fine-tuning gain is that of the whole dependency-guided procedure, since no alternative adaptation objective was evaluated, so it is not attributed to dependency topology alone.

\section{Conclusion\label{sec:conclusion}}

This study examines how code text, static dependencies, folder organization, and member-level information contribute to code-to-architecture mapping on ten systems. The findings show that mapping performance depends not only on which information is available but also on how it is represented and used.

Representations that encode dependency information generally outperform the evaluated lexical and zero-shot code representations. Node2Vec and dependency-guided fine-tuned UniXcoder outperform Word2Vec on most systems, and fine-tuning improves UniXcoder over its zero-shot counterpart on nearly all systems. However, neither Node2Vec nor fine-tuned UniXcoder is consistently best across systems.

Folder nodes produce the largest and most consistent gains within the HGT variants, but a directory rule using the same labeled files reaches similar accuracy on most systems, and folder paths supplied as file features match or exceed folder nodes on most systems. These gains do not consistently translate into better predictions on files the rule assigns incorrectly. Member nodes provide no consistent improvement with one or two HGT layers, and their effects cannot generally be separated from changes in feature construction and placement. These structural findings are specific to HGT; whether they persist with other predictors remains untested.

Keeping the input representations fixed while enabling and disabling access to dependency-neighborhood information shows that neighbor aggregation can complement representations that already encode dependencies. With undirected aggregation, the median change across systems is positive on both metrics for every predictor and representation. In this setting, GAT achieves the highest median macro-F1, while HGT gains the least. The better aggregation direction varies between systems, although the experiments do not isolate propagation direction from neighborhood size. Evaluation should therefore distinguish information already encoded in the representations from the contribution of neighborhood access. The findings support assessing these contributions for the intended mapping task, rather than treating any representation or predictor as generally preferable.

\parg{Future work.} Future research could clarify how different aspects of graph context contribute to the observed differences. Comparing incoming, outgoing, and combined neighborhoods at matched sizes would help distinguish the effects of propagation direction and neighborhood size. Adding member nodes while preserving each file's own features would help separate the effects of graph structure and feature placement. Another direction is configuration selection on unfamiliar systems. Homophily, which helped interpret the system differences, requires reference module assignments, so selection rules based on properties observable in the dependency graph alone could be developed on some systems and evaluated on others. Finally, a model trained on one version could be evaluated on new and changed files in later versions without retraining, to test whether the learned information supports mapping as systems evolve.

\bibliographystyle{ACM-Reference-Format}
\bibliography{References}

\end{document}